\newif\ifdoubleblind
\newif\ifacm
	
\ifacm
	\documentclass[sigconf]{acmart}
\else
	\documentclass[conference]{IEEEtran}
\fi

\usepackage{amsmath}

\usepackage{multirow}
\usepackage{tabularx,booktabs}
\usepackage{xspace}

\usepackage[binary-units=true]{siunitx}
\usepackage{tikz}
\usetikzlibrary{positioning}

\ifacm
	\usepackage[nolist]{acronym}
\usepackage{booktabs} 
\usepackage{subfig}
\usepackage{balance}

\usepackage{array}
\usepackage{graphicx}
\usepackage{wasysym}
\usepackage{xparse}

\usepackage{makecell}
\newcolumntype{Y}{>{\centering\arraybackslash}X}

\hypersetup{draft}

\renewcommand\footnotetextcopyrightpermission[1]{} 
\else
	\usepackage{graphicx}
\usepackage{epstopdf}
\usepackage{standalone}
\usepackage[nolist ]{acronym}
\usepackage{tcolorbox}
\usepackage{pdfpages}
\usepackage[bookmarks=false]{hyperref}
\usepackage{pdfcomment}
\usepackage{hologo}

\usepackage{xstring}

\usepackage[utf8]{inputenc}
\usepackage{marvosym}

\usepackage{algorithm}
\usepackage{algpseudocode}

\usepackage{psfrag}
\usepackage{graphicx}

\usepackage{rotating}

\renewcommand{\baselinestretch}{1}

\usepackage[caption=false,font=footnotesize]{subfig}
\usepackage{stfloats}
\fi
\graphicspath{fig}
\begin{document}

\newcommand{\paperTitle}{Towards Machine Learning-Enabled Context Adaption for Reliable Aerial Mesh Routing}

\newcommand{\paperAuthors}{Cedrik Schüler, Benjamin Sliwa, and Christian Wietfeld}
\newcommand{\paperEmails}{$\{$Cedrik.Schueler, Benjamin.Sliwa, Christian.Wietfeld$\}$@tu-dortmund.de}

\newcommand\single{1\textwidth}
\newcommand\double{.48\textwidth}
\newcommand\triple{.32\textwidth}
\newcommand\quarter{.24\textwidth}
\newcommand\singleC{1\columnwidth}
\newcommand\doubleC{.475\columnwidth}

\newcommand{\figurePadding}{0pt}
\newcommand{\figureTopPadding}{\figurePadding}
\newcommand{\figureBottomPadding}{\figurePadding}

\newcommand\err[1]{\colorbox{red}{\textbf{#1}}}
\newcommand\wrn[1]{\colorbox{orange}{\textbf{#1}}}
\newcommand\grn[1]{\colorbox{green}{\textbf{#1}}}

\newcommand\rtxt[1]{\textcolor{red}{#1}}
\newcommand\ytxt[1]{\textcolor{orange}{#1}}
\newcommand\gtxt[1]{\textcolor{olive}{#1}}

\newcommand\tikzFig[2]
{
	\begin{tikzpicture}
		\node[draw,minimum height=#2,minimum width=\columnwidth,text width=\columnwidth,pos=0.5]{\LARGE #1};
	\end{tikzpicture}
}

\newcommand{\dummy}[3]
{
	\begin{figure}[b!]  
		\begin{tikzpicture}
		\node[draw,minimum height=6cm,minimum width=\columnwidth,text width=\columnwidth,pos=0.5]{\LARGE #1};
		\end{tikzpicture}
		\caption{#2}
		\label{#3}
	\end{figure}
}

\newcommand\pos{h!tb}

\newcommand{\basicFig}[7]
{
	\begin{figure}[#1]  	
		\vspace{#6}
		\centering		  
		\includegraphics[width=#7\columnwidth]{#2}
		\caption{#3}
		\label{#4}
		\vspace{#5}	
	\end{figure}
}
\newcommand{\fig}[4]{\basicFig{#1}{#2}{#3}{#4}{0cm}{0cm}{1}}

\newcommand\sFig[2]{\begin{subfigure}{#2}\includegraphics[width=\textwidth]{#1}\caption{}\end{subfigure}}
\newcommand\vs{\vspace{-0.3cm}}
\newcommand\vsF{\vspace{-0.4cm}}

\newcommand{\subfig}[3]
{%
	\subfloat[#3]%
	{%
		\includegraphics[width=#2\textwidth]{#1}%
	}%
	\hfill%
}

\newcommand\circled[1] 
{
	\tikz[baseline=(char.base)]
	{
		\node[shape=circle,draw,inner sep=1pt] (char) {#1};
	}\xspace
}
\begin{acronym}
	\acro{B.A.T.M.A.N.}{Better Approach To Mobile Ad-hoc Networking}
	\acro{OMNeT++}{Objective Modular Testbed in C++}
	\acro{LIMITS}{LIghtweight Machine learning for IoT Systems}
	\acro{LIMoSim}{Lightweight ICT-centric Mobility Simulation}
	\acro{MANET}{Mobile Ad-hoc Network}
	\acro{VANET}{Vehicular Ad-hoc Network}
	\acro{FANET}{Flying Ad-hoc Network}
	\acro{UAV}{Unmanned Aerial Vehicle}
	\acro{ITS}{Intelligent Transportation System}
	\acro{AODV}{Ad-hoc On-demand Distance Vector}
	\acro{DYMO}{Dynamic MANET On Demand Routing Protocol}
	\acro{OLSR}{Optimized Link State Routing}
	\acro{GPSR}{Greedy Perimeter Stateless Routing in Wireless Networks}
	\acro{DSDV}{Destination-Sequenced Distance Vector}
	\acro{PARRoT}{Predictive Ad-hoc Routing fueled by Reinforcement learning and Trajectory knowledge}
	\acro{PDR}{Packet Delivery Ratio}
	\acro{UDP}{User Datagram Protocol}
	\acro{CBR}{Constant Bitrate}
	\acro{LOS}{Line-of-Sight}
	\acro{RREQ}{Route Request}
	\acro{RREP}{Route Reply}
	\acro{LET}{Link Expiry Time}
	\acro{LIMoSim}{LIghtweight ICT-centric Mobility Simulation}
	\acro{MCN}{Multi-hop Cellular Network}
	\acro{ML}{Machine Learning}
	\acro{RL}{Reinforcement Learning}
	\acro{MPR}{Multipoint Relay}
	\acro{CRS-MP}{Centralized Routing Scheme with Mobility Prediction}
	\acro{SDN}{Software-defined Networking}
	\acro{ANN}{Artificial Neural Network}
	\acro{DDD}{Distributed Dispersion Detection}
	\acro{TTL}{Time to Live}
	\acro{SEQ}{Sequence Number}
	\acro{HWMP}{Hybrid Wireless Mesh Protocol}
	\acro{MAC}{Medium Access Control}
	\acro{GPS}{Global Positioning System}
	\acro{ANN}{Artificial Neural Network}
	\acro{SVM}{Support Vector Machine}
	\acro{RF}{Random Forest}
	\acro{GULAG}[CA-PARRoT]{Context-Adaptive PARRoT}
	\acro{KPI}{Key Performance Indicator}
	\acro{RSS}{Received Signal Strength}
	\acro{REP}{Radio Environment Prototype}
	\acro{SMO}{Sequential Minimal Optimization}
\end{acronym}

\title{\paperTitle}

\ifacm
\else

	\title{\paperTitle}

	\ifdoubleblind
	\author{\IEEEauthorblockN{\textbf{Anonymous Authors}}
		\IEEEauthorblockA{Anonymous Institutions\\
			e-mail: Anonymous Emails}}
	\else
	\author{\IEEEauthorblockN{\textbf{\paperAuthors}}
		\IEEEauthorblockA{Communication Networks Institute,	TU Dortmund University, 44227 Dortmund, Germany\\
			e-mail: \paperEmails}}
	\fi
	
	\maketitle

\fi




\begin{abstract}
	%
	%
In this paper, we present \ac{GULAG} as an extension of our previous work \ac{PARRoT}.
Short-term effects, as occurring in urban surroundings, have shown to have a negative impact on the \ac{RL}-based routing process. Therefore, we add a timer-based compensation mechanism to the update process and introduce a hybrid \ac{ML} approach to classify \acp{REP} with a dedicated \ac{ML} component and enable the protocol for autonomous context adaption.
%
%
The performance of the novel protocol is evaluated in comprehensive network simulations considering different \acp{REP} and is compared to well-known established routing protocols for \acp{MANET}.
The results show, that \ac{GULAG} is capable to compensate the challenges confronted with in different \acp{REP} and to improve its \acp{KPI} up to  23\,\% compared to \ac{PARRoT}, and outperform established routing protocols by up to 50\,\%. 

\end{abstract}

\maketitle

\begin{tikzpicture}[remember picture, overlay]
\node[below=5mm of current page.north, text width=20cm,font=\sffamily\footnotesize,align=center] {Accepted for presentation in: 2021 IEEE 94rd Vehicular Technology Conference (VTC-Fall)\vspace{0.3cm}\\\pdfcomment[color=yellow,icon=Note]{
@InProceedings\{Schueler/etal/2021b,\\
  author    = \{Cedrik Sch\{\"u\}ler and Benjamin Sliwa and Christian Wietfeld\},\\
  booktitle = \{2021 IEEE 94rd Vehicular Technology Conference (VTC-Fall)\},\\
  title     = \{Towards machine learning-enabled context adaption for reliable aerial mesh routing\},\\
  year      = \{2021\},\\
  address   = \{Virtual\},\\
  month     = \{Sep\},\\
\}
}};
\node[above=5mm of current page.south, text width=15cm,font=\sffamily\footnotesize] {2021~IEEE. Personal use of this material is permitted. Permission from IEEE must be obtained for all other uses, including reprinting/republishing this material for advertising or promotional purposes, collecting new collected works for resale or redistribution to servers or lists, or reuse of any copyrighted component of this work in other works.};
\end{tikzpicture}

\section{Introduction}

%
%
Aerial mesh networks offer the ability to be deployed spontaneously for events or by emergency personnel to manage desaster situations. Although this enables wide application ranges, certain challenges arise throughout these scenarios.

\begin{figure}[]  	
	\vspace{0cm}
	\centering		  
	\includegraphics[width=1.0\columnwidth]{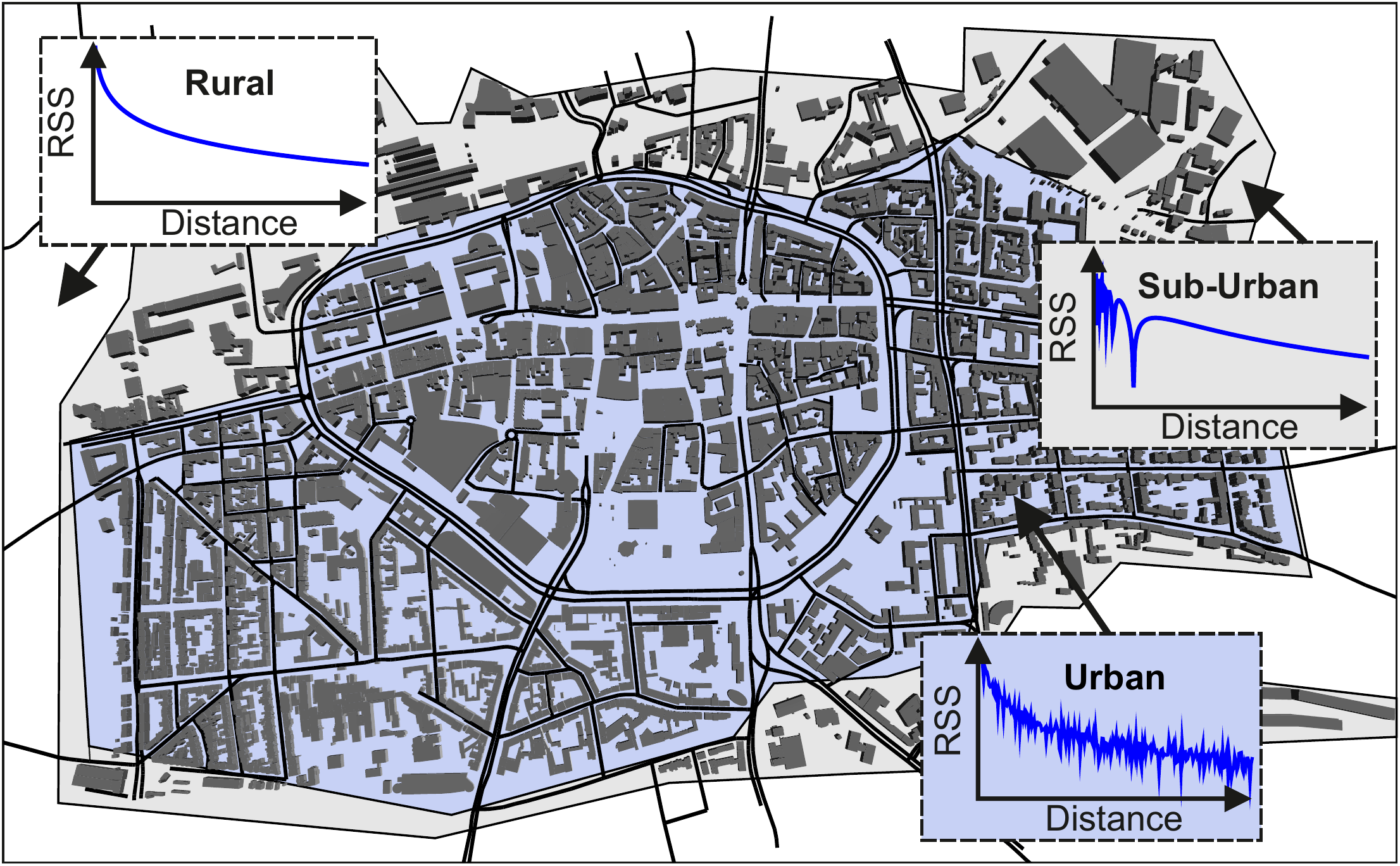}
	\caption{Example of an application scenario, consisting of multiple radio environment prototypes (Map data: $\copyright$ OpenStreetMap Contributors, CC BY-SA).}
	\label{fig:scenario}
	\vspace{0cm}	
\end{figure}

Besides high agent's mobilities and hybrid vehicle types participating in the network, the agents must also be able to operate in different \acp{REP}.
Multi-\ac{REP} scenarios (see Fig.~\ref{fig:scenario}) are considerably the most challenging situations for autonomous aerial mesh networks, as the mission cannot be interrupted to reconfigure the drones to fit a new \ac{REP}, especially if they have a high mobility and \ac{REP} changes are frequent.
However, even if the application is limited to a single-\ac{REP} area, an autonomous context adaption is neccessary to enable a quick network deployment without pre-evaluation.


Our previous work \cite{Sliwa/etal/2021a} has shown, that mobility-predictive routing protocols, such as our proposed \ac{PARRoT}, and B.A.T.Mobile \cite{Sliwa/etal/2016a} outperform the considered well-known protocols in terms of robustness and reliability within different stress tests, channel conditions, and agent mobilities.
Although, the reinforcement learning-based \ac{PARRoT} widely performed best, a high impact of short-term channel variations on the \ac{PDR} has been observed and motivated further enhancement.
Thus, we present \ac{GULAG}, extending \ac{PARRoT} with a compensation mechanism and introducing a hybrid \ac{ML} approach.
%
%

The contributions are summarized as follows:
%
%
\begin{itemize}
	\item Presentation of the enhanced \textbf{\ac{GULAG}} protocol and extension of our open-source C++ simulation framework for \ac{OMNeT++} \cite{Varga/Hornig/2008a}.\footnote{The simulation framework is available under: \url{https://www.github.com/cedrikschueler/PARRoT}}
	\item Parameterization study for different \acp{REP} and proposal on a \textbf{machine learning-based parameter selection}.
	\item \textbf{Performance analysis} of \ac{GULAG} in comparison to well-established routing protocols, as well to our mobility-predictive representatives B.A.T.Mobile and \ac{PARRoT}.
\end{itemize}

%
%
The remainder of the paper is structured as follows. After discussing the related work in Sec.~\ref{sec:related_work}, we present the extensions made to the \ac{GULAG} protocol, and an approach to auto-adaptive parameterization in Sec.~\ref{sec:approach}. Afterwards, an overview about the methodological aspects is given in Sec.~\ref{sec:methods}. Finally, detailed results are provided and discussed in Sec.~\ref{sec:results}.

%
%

\section{Related Work} \label{sec:related_work}
%
%
\ac{UAV} mesh networks are deployed in recent, diverse applications, such as modern logistics, to provide contactless delivery systems or disinfect larger areas during COVID-19 pandemic \cite{Chamola/etal/2020a}.
Furthermore, the air-assisted traffic control in \acp{ITS} allows the exchange of safety-critical messages, but, also infotainment related data among and across hybrid vehicular types, to enable autonomous driving and better traffic efficiency in smart cities \cite{Menouar/etal/2017a}\cite{Zeng/etal/2019a}.

%
%
The empirical analysis of Cavalcanti et al. \cite{Cavalcanti/etal/2018a} points out the popularity of routing protocols, channel models, and simulation methods for vehicular communication research.
A summary of \ac{MANET} routing protocols is given in \cite{Oubbati/etal/2019a}. A common distinction is the classification into \textit{reactive} protocols --e.g. \ac{AODV} and \ac{DYMO}--, that build up routes on demand, and \textit{proactive} protocols, which maintain routes by periodically broadcasting their information. Well-known representatives of the latter are \ac{OLSR}, \ac{DSDV}, and \ac{B.A.T.M.A.N.}.

Further approaches, like \ac{GPSR}, include \textit{geo-based} knowledge, like position and velocity, into the routing process, to respect the impact of mobility on communication capabilities. \ac{PARRoT} \cite{Sliwa/etal/2021a} and B.A.T.Mobile \cite{Sliwa/etal/2016a} are considered \textit{geo-predictive}, as they utilize cross-layer mobility knowledge to predict agent's positions and make their routing decisions.
%
%
Thus, they follow the anticipatory mobile networking \cite{Bui/etal/2017a} paradigm, which describes the information aggregation to predict the network situation and not being forced to react, but to be prepared for upcoming changes.
Besides mobility prediction, to assess topology changes, context-classification is an important part of anticipatory networking.
%
%
Machine learning techniques have attracted a keen interest of the wireless research community due to their inherent of solving prediction and optimization problems in complex environments  \cite{Wang/etal/2020a}. As a result, they are expected to be an essential part of upcoming network generations \cite{Ali/etal/2020a}. In this paper, we utilize \ac{ML} methods to compete the challenging task of channel classification \cite{viriyasitavat/etal/2015}, by learning from channel parameters \cite{bharti/etal/2020}.

\section{Proposed Extensions to the PARRoT Routing Protocol} \label{sec:approach}
\begin{figure}[]  	
	\vspace{0cm}
	\centering		  
	\includegraphics[width=1.0\columnwidth]{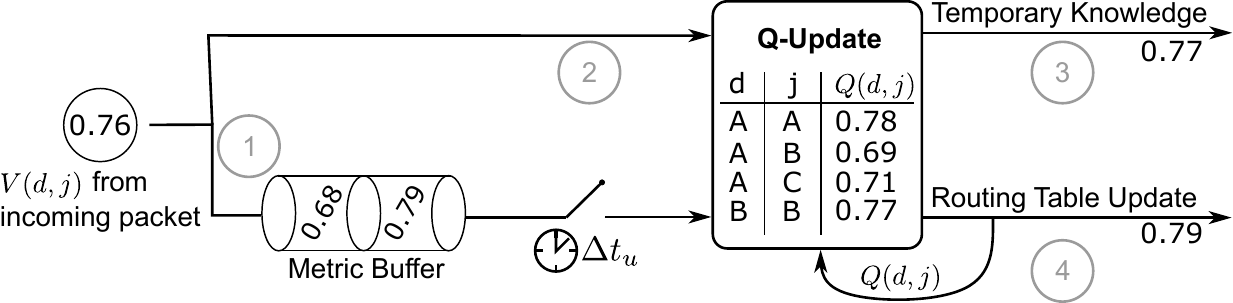}
	\vspace{-0cm}	
	\caption{Architecture of the timer-based knowledge management in \ac{PARRoT}.}
	\label{fig:timer_principle}
	\vspace{-0cm}	
\end{figure}
In this section, we present the proposed adaptions made to \ac{GULAG}, after briefly discussing the fundamentals of the underlying \ac{PARRoT} routing protocol, which we refer to our previous work \cite{Sliwa/etal/2021a}.
Afterwards, we introduce additional parameters to the \ac{RL} and present a \ac{ML}-based component, dedicated to classify the conditions and select appropiate parameters to enable a zero-touch deployment.
%
%
\subsection{\ac{PARRoT} Fundamentals}\label{sec:approach:parrot_fundamentals}
\ac{PARRoT} joins the core ideas of integrating mobility domain knowledge to predict agent trajectories, and derive autonomous routing decisions, based on agents' relations and link availabilities.
\ac{PARRoT} inherits the mobility prediction from B.A.T.Mobile~\cite{Sliwa/etal/2016a}. Every agent approaches to predict its position $\tilde{\mathbf{p}}(t + \tau)$ for a future timestep with a prediction width $\tau$.
For this purpose, $\tau$ is divided into equidistant timesteps on which an iterative process is applied. Based on the temporary predicted position, the trajectory is moved towards the next target, if available.
If no more target points can be retrieved from the mobility domain, remaining prediction steps are extrapolated from the position history.

All agents periodically broadcast routing messages, which wrap identification data, metrics, and position information in a total of 40\,Bytes. As originator, the contained reward metric is set to a value of $1$.
Upon a received message, neighbor information is extracted and the agent invokes the reinforcement learning update process, and an update of its routing table.
If the agent can contribute to the route building process, it emplaces its mobility information and updates the reward metric to its own best estimation for a route to the originator. The message is then forwarded and propagates through the network.
By this procedure, every agent generates a Q-table with possibly multiple entries to reach a destination node. By default, the routing behaves always greedy and the next hop with the best metric score is selected. Neighbors can be considered as forwarders for the time scope of the predicted link availability or a maximum of $\tau$.

Aforementioned link availability is calculated based on the position $\mathbf{p}(t)$ and the predicted position $\tilde{\mathbf{p}}(t + \tau)$ of the current node $i$, and a candidating neighbor node $j$, respectively. The relative trajectories are compared with respect to the maximum communication range $r_{\textrm{TX}}$, which is derived from a freespace model. As an output, the time where $i$ and $j$ can communicate is further processed into the \ac{RL} metric $\Phi_{\textrm{LET}}(i,j)$ and is also used to track the validity time of entries.
\subsection{Variance-compensating Update Process}\label{sec:approach:updateprocess}
Radio signals propagate via multiple paths in urban environments, causing constructive or destructive interference at the receiver. The wireless channel is therefore affected by influences that may only be valid for a short period of time. Especially in the routing context, this leads to incorrect assessments of the current situation and can be a main reason for performance losses.
To overcome this issue, a measure is neccessary, to prevent occasionally occuring input from foreign agents from having too much impact on the reinforcement learning-based routing process and the stored knowledge. On the other side, missed out messages from neighboring nodes need to be handled.

Within our previous B.A.T.Mobile~\cite{Sliwa/etal/2016a} protocol, a timer-based approach was introduced addressing the described problem. Incoming routing packages only update a current score candidate $S_C$, which is shifted to a \textit{neighbor score buffer} after an elapsed update interval $t_u$.
Concerning the update process in \ac{GULAG}, a small adaption has to be made:

Incoming chirps contain the best effort estimation $V(d, j)$ of the last forwarder of the message $j$, to reach a target destination $d$. The forwarding process utilizes the most recent metric by calculating the quality indicator $Q(d, j)$, and, then decides to forward the chirp or not.
In \ac{PARRoT}, this metric was stored upon every reception of a chirp.
Because the Q-learning is implemented as an online learning process, this behaviour leads to an very unstable Q-table, that is easily affectable by incoming packets.

\ac{GULAG} stores route information depending on the freshness (\textit{is the sequence number higher?}), and its worth (\textit{is $V$ higher than the stored one?}).
This means, that only the \textit{best candidates} are stored and other information is discarded.
To reduce the variance of stored knowledge, we supply \ac{GULAG} with a timer-based update as shown in Fig.~\ref{fig:timer_principle}.

The new update routine is divided into four parts and works as follows:
\begin{enumerate}
	\item Neighbor-related information is always stored (i.e. position, velocity, $\Phi_{\textrm{Coh}}(j)$), but route information is only stored in the metric buffer if \begin{itemize}
		\item The information is more recent
		\item The received metric is higher
		\end{itemize}
	\item The recently buffered value triggers a calculation of the Q-value, for which, the currently stored value  $Q(d, j)$ is queried. This leads to,
	\item A temporary valid knowledge, with which the further processing of the chirp is determined.
	\item Triggered by a periodic timer, the buffered knowledge is overwritten with the most recent route candidate. Also, this timer incorporates an update of the routing tables.
\end{enumerate}

\subsection{Machine Learning-enabled Estimation of the Channel Properties}
\begin{figure}[]  	
	\vspace{0cm}
	\centering		  
	\includegraphics[width=1.0\columnwidth]{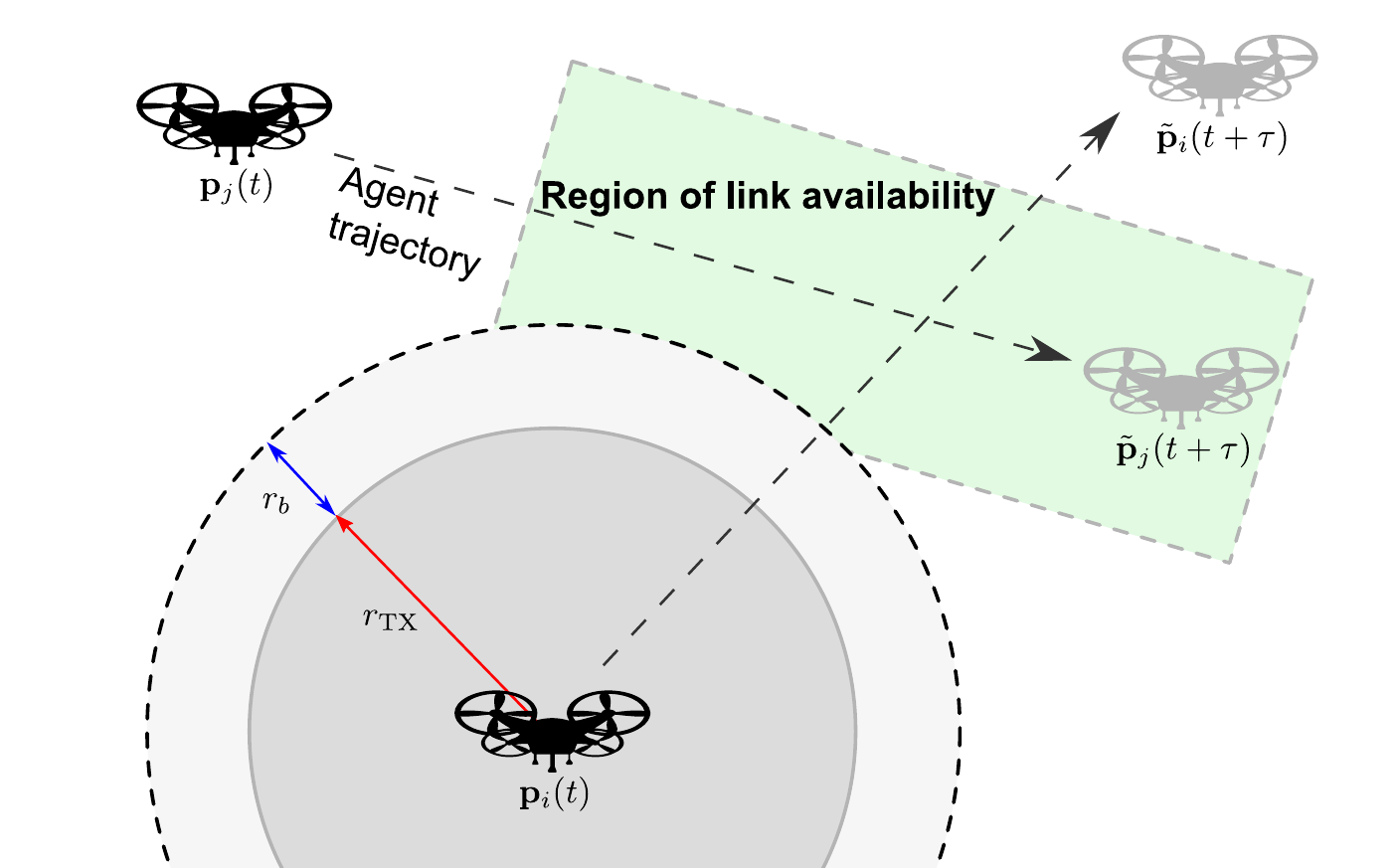}
	\vspace{-0cm}	
	\caption{\ac{GULAG}'s approach to assess link availabilities between two agents $i$ and $j$ by utilizing agent trajectories to predicted positions $\tilde{\mathbf{p}}(t+\tau)$.}
	\label{fig:parrot_illustration}
	\vspace{-0cm}	
	\end{figure}
Routing protocols need to handle different circumstances, which cannot be fulfilled  with a fixed parameterization.
In order to make \ac{GULAG} more adoptable for disjunct \acp{REP}, we consider the range budget $r_b$ as a \ac{REP} specific parameter.
As described in Sec.~\ref{sec:approach:parrot_fundamentals} a assumed communication range $r_{\textrm{TX}}$ is derived from a freespace channel model, and is used in the calculation of the partial metric $\Phi_{\textrm{LET}}$.
The range budget $r_b$ is added to this value, leading the edge condition for the relative agent trajectories $\Delta\tilde{\mathbf{p}} = \Delta\mathbf{p} + t\cdot\Delta\mathbf{v}$, with the relative agent position $\Delta\mathbf{p}$, and velocity $\Delta\mathbf{v}$ to be
\begin{equation}
	r_{\textrm{TX}} + r_b = \sqrt{\Delta\tilde{\mathbf{p}}_x^2 + \Delta\tilde{\mathbf{p}}_y^2 + \Delta\tilde{\mathbf{p}}_z^2}.
\end{equation}
This then resolves to the metric $\Phi_{\textrm{LET}}(i, j)$ between the current agent $i$ and a neighbor agent $j$ in accordance with \cite{Sliwa/etal/2021a}.
Fig.~\ref{fig:parrot_illustration} illustrates the impact of $r_b$ on the \ac{LET} calculation.

Additionally, the partial metrics can be of different importance in other \acp{REP}. To address this, we modify the Q-learning formula, and add a link weighting $\lambda$ for $\Phi_{\textrm{LET}}(i, j)$ and a cohesion weighting $\omega$ for $\Phi_{\textrm{Coh}}(j)$ respectively.
The basic discount factor $\gamma_0$ still guarantees a path degragation for $\gamma_0 \in [0, 1)$.
Thus, the Q-learning update formula defines as
\begin{equation}
	\begin{split}
		Q(d, j) &= Q(d, j) + \alpha[\gamma(j) \cdot V(d, j) - Q(d, j)] \\
		\text{with: }\gamma(j) &= \gamma_0\cdot \Phi_{\textrm{LET}}^\lambda(i,j) \cdot \Phi_{\textrm{Coh}}^\omega(j).
	\end{split}
	\label{equ:weighted_qlearning}
\end{equation} 

\begin{figure}[]  	
	\vspace{0cm}
	\centering		  
	\includegraphics[width=1.0\columnwidth]{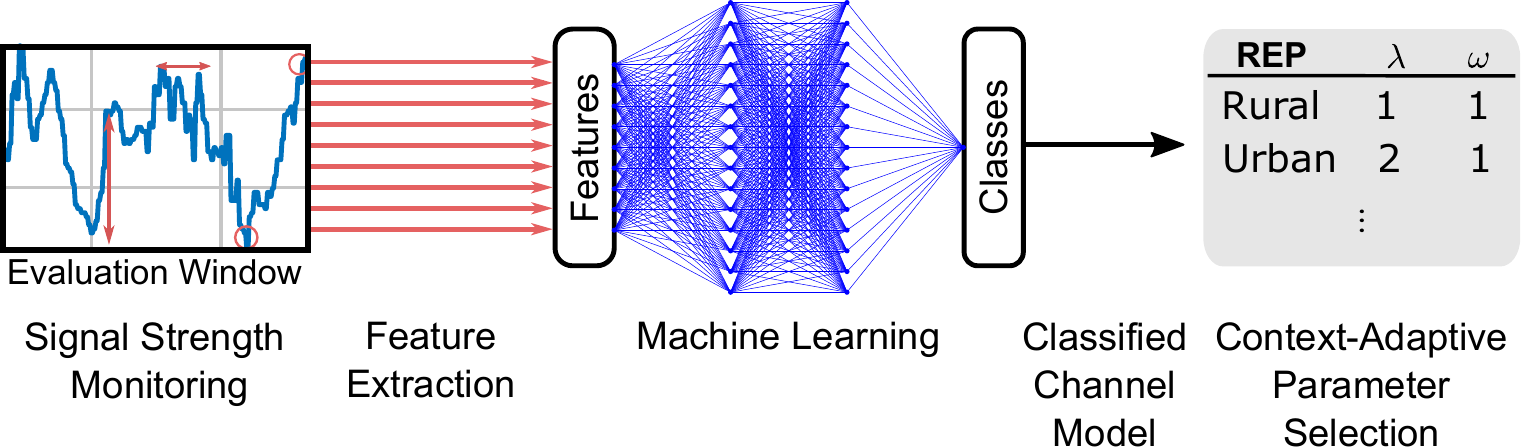}
	\vspace{-0cm}	
	\caption{Concept of the \ac{ML}-based, context-adaptive parameter selection.}
	\label{fig:self_adaption}
	\vspace{-0cm}	
\end{figure}

The proposed extensions lead to a parameterization tuple of $(r_b, \alpha, \gamma_0, \lambda, \omega)$ which is pre-evaluated for different \acp{REP} and is stored in a database.
As shown in Fig.~\ref{fig:self_adaption}, each agent monitors the \ac{RSS} and distance pairs for incoming routing messages, from which features are extracted and used to classify a \ac{REP} by a \ac{ML} component. Based on the result, a parameter set is selected.

\begin{figure}[]  	
	\vspace{0cm}
	\centering		  
	\includegraphics[width=1.0\columnwidth]{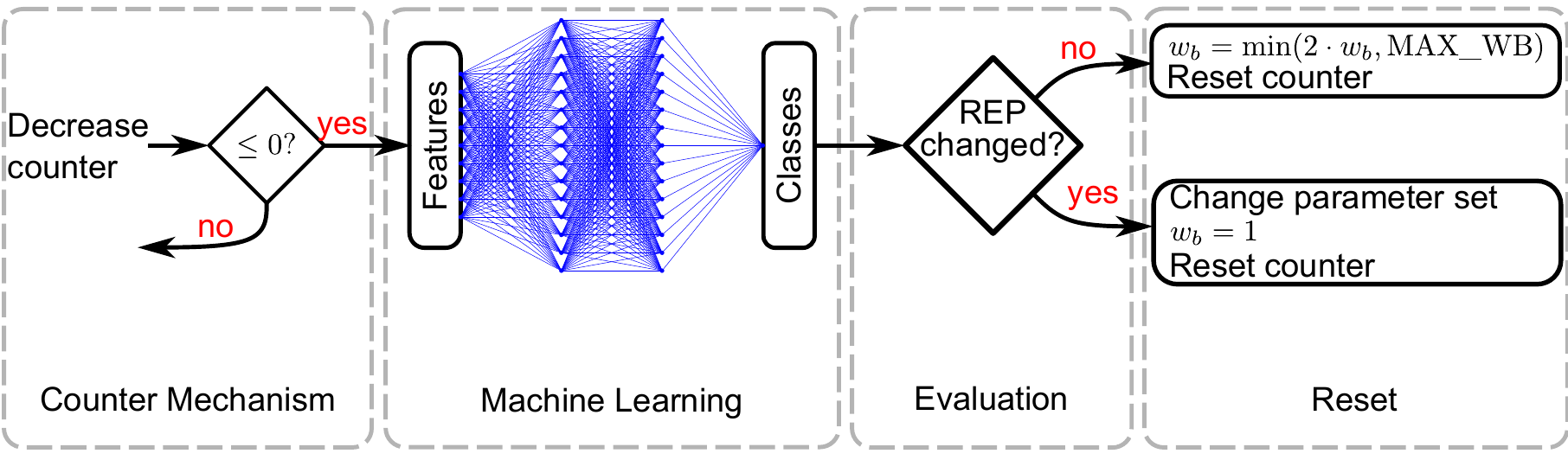}
	\vspace{-0cm}	
	\caption{Workflow of the backoff counter and the modification of the backoff window $w_b$.}
	\label{fig:backoff_counter}
	\vspace{-0cm}	
\end{figure}

To reduce the amount of re-calculations in stable environments, a backoff counter is implemented. Environment checks always take place before the Q-table update as described in Sec.~\ref{sec:approach:updateprocess}. The backoff counter is initially set to the length of the backoff window $w_b$ and decreases with every timer event.
The environment check is performed, if the counter reaches a value of $0$ and skipped otherwise.
Fig.~\ref{fig:backoff_counter} indicates the following condition: If the \ac{REP} changed, $w_b$ is reset to $1$ to ensure a fast validation of the new estimation. If the previous classification is confirmed, $w_b$ grows exponentially and, therefore, increases the trust in the current prototype classification. Regardless of how $w_b$ changes, the backoff counter needs to be reset to the new value afterwards.
\section{Methodology} \label{sec:methods}
In this section, we present the methodological aspects of the performance evaluation in Sec.~\ref{sec:results}.
All simulations are carried out with the \ac{OMNeT++}~\cite{Varga/Hornig/2008a} simulator, using \ac{MANET} protocol implementations from the \textit{INETMANET} framework.
A \acs{UDP} \ac{CBR} video stream between two hosts is established to gather the \ac{PDR} and latency as end-to-end \acp{KPI}.
The agents move within a three-dimensional playground, following a controlled waypoint pattern.This is a variation of the well-known random waypoint mobility, with the extension, that future waypoints are pre-calculated and, thus, can be exploited by mobility prediction.

\textbf{\acfp{REP}:} To evaluate the proposed context-aware self adaption, we consider three \acp{REP}, \acp{MANET} are possibly deployed in.
\begin{itemize}
	\item \textbf{Rural} environments provide fairly simple radio conditions, where pathloss is primarly caused by signal attenuation over distance.
	\item \textbf{Sub-urban} environments contain scattered objects, that cause an additional reflecting path to the line-of-sight component in near-field situations.
	\item \textbf{Urban} areas are characterized by complex surroundings, leading to multipath propagations, interfering with the \ac{LOS} path, and, thus, creating a highly challenging radio prototype.
\end{itemize} 
%
%
\begin{table}[ht]
	\centering
	\caption{Default parameters of the evaluation setup}
	\begin{tabular}{lr}
		\toprule
		\textbf{Parameter} & \textbf{Value} \\

		\midrule
		
		\ac{OMNeT++} version & 5.6.1 \\
		INETMANET version & 4.x \\
		MAC & 802.11g \\
		Rural channel model & Friis ($\eta= 2.75$) \\
		Sub-urban channel model & Two-Ray Ground \\ 
		Urban channel model & Nakagami ($\eta= 2.75$, $m=2$) \\
		Playground size & 500\,m x 500\,m x 250\,m \\
		Number of runs per configuration & 25 \\
		Simulation time & 900\,s \\
		Number of routing hosts & 10 \\
		
		Transmission power & 20\,dBm \\
		Receiver sensitivity & -85\,dBm \\
		
		Mobility model & Controlled waypoint \\

		Speed & 50\,km/h \\
		Traffic load per video stream & 2\,Mbit/s \\
		Chirp interval $\Delta t_{\textrm{Chirp}}$ & \SI{0.5}{s} \\
		\bottomrule
		
	\end{tabular}
	\label{tab:parameters}
\end{table}

%
%
Tab.~\ref{tab:parameters} summarizes the simulation parameters. The respective channel models were chosen due to their popularity in vehicular research according to \cite{Cavalcanti/etal/2018a}. As well, due to their popularity, we consider 
\begin{itemize}
	\item \textbf{\ac{AODV}} as a \textit{reactive} distance vector routing protocol,
	\item \textbf{\ac{OLSR}} as a \textit{proactive} link state approach,
	\item \textbf{\ac{GPSR}} as a \textit{geo-based} distance minimization protocol,
	\item \textbf{B.A.T.Mobile} and \textbf{\ac{PARRoT}} as our \textit{geo-predictive} previous works,
\end{itemize}
as references for the performance evaluation.

The training and evaluation of the \ac{REP} classification are performed with the \textit{LIMITS}~\cite{Sliwa/etal/2020c} framework. For the \ac{REP} classification task, we consider the following machine learning models whereas the parameters are chosen based on an initial grid search optimization:

\begin{itemize}
	\item \textbf{\acf{ANN}} \cite{LeCun/etal/2015a} with two hidden layers, 10 neurons per hidden layer, learning rate $\eta=0.1$, momentum $\alpha=0.01$, sigmoid activation function, and 500 training epochs.
	\item \textbf{\acf{RF}} \cite{Breiman/2001a} using 100 random trees and a maximum depth of 15.
	\item Linear \textbf{\acf{SVM}} \cite{Cortes/Vapnik/1995a} trained via \ac{SMO}.
\end{itemize}

In the later result analysis, we analyze the 10-fold cross validation accuracy of each model. The best performing classification model is then embedded into \ac{GULAG} for performing the online classifications during the simulation runs.

\begin{figure}[]  	
	\vspace{0cm}
	\centering		  
	\includegraphics[width=0.75\columnwidth]{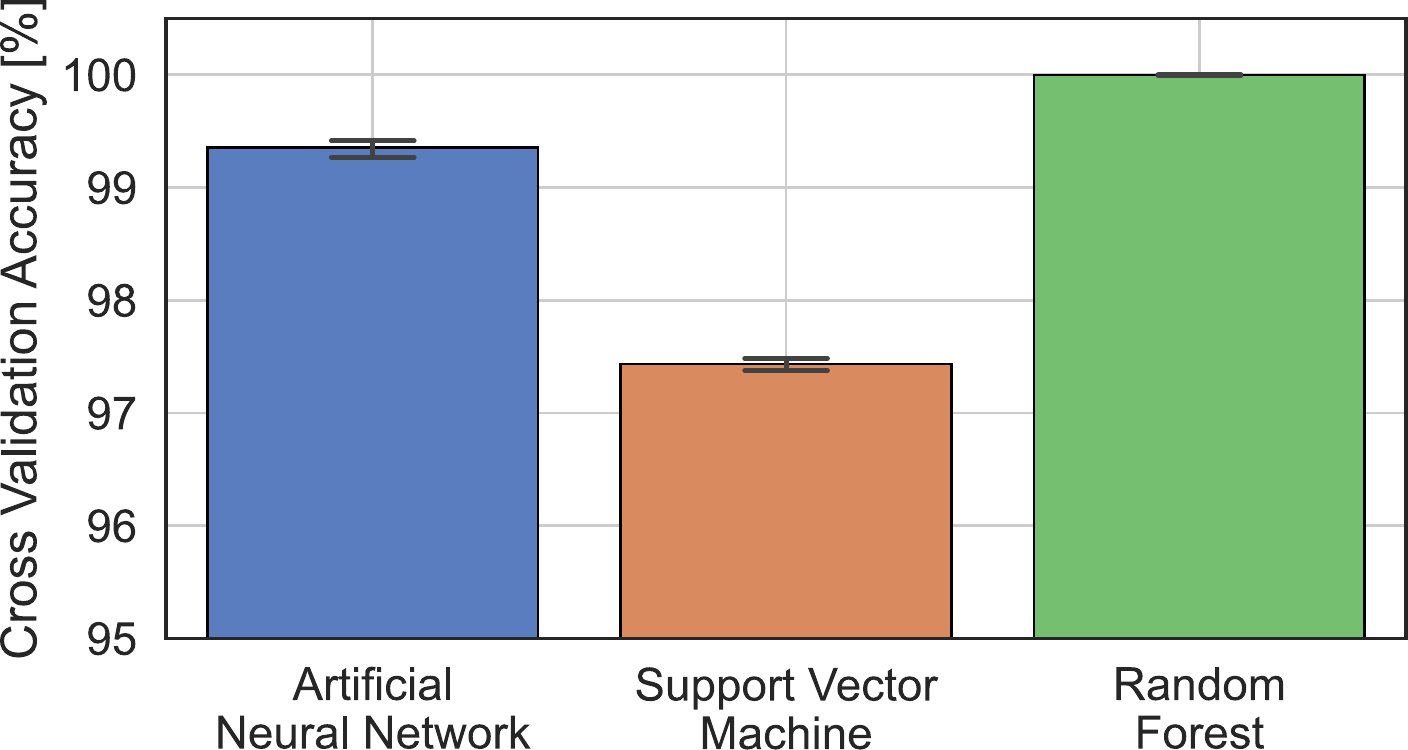}
	\vspace{-0cm}	
	\caption{Cross validation accuracy of different \ac{ML} classification approaches.}
	\label{fig:ml_accuracy}
	\vspace{-0cm}	
\end{figure}

\begin{figure}[]  	
	\vspace{0cm}
	\centering		  
	\includegraphics[width=1.0\columnwidth]{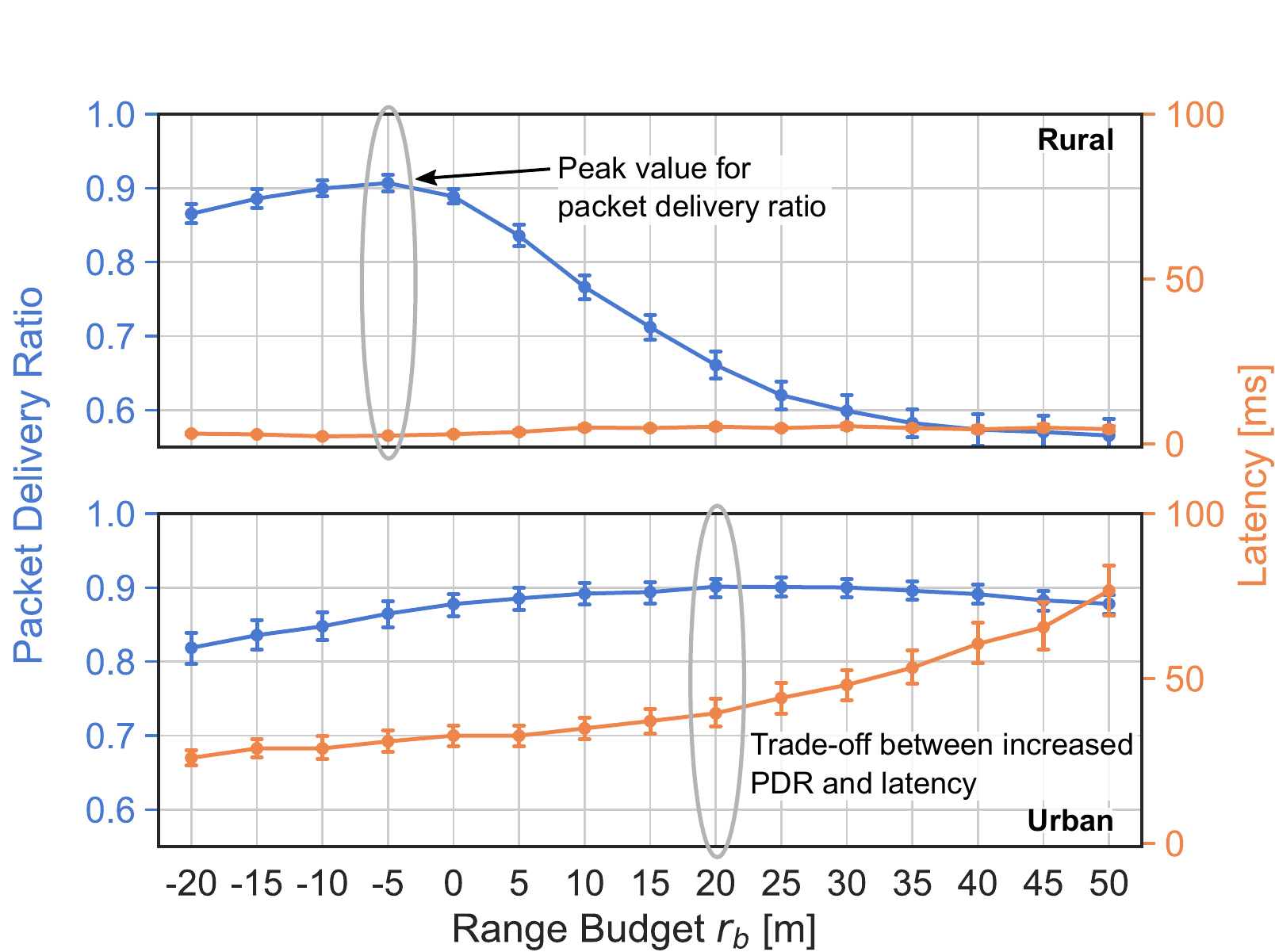}
	\vspace{-0cm}	
	\caption{Range budgets $r_b$ for rural and urban radio prototypes.}
	\label{fig:range_budget}
	\vspace{-0cm}	
\end{figure}

\begin{figure}[]  	
	\vspace{0cm}
	\centering		  
	\includegraphics[width=1.0\columnwidth]{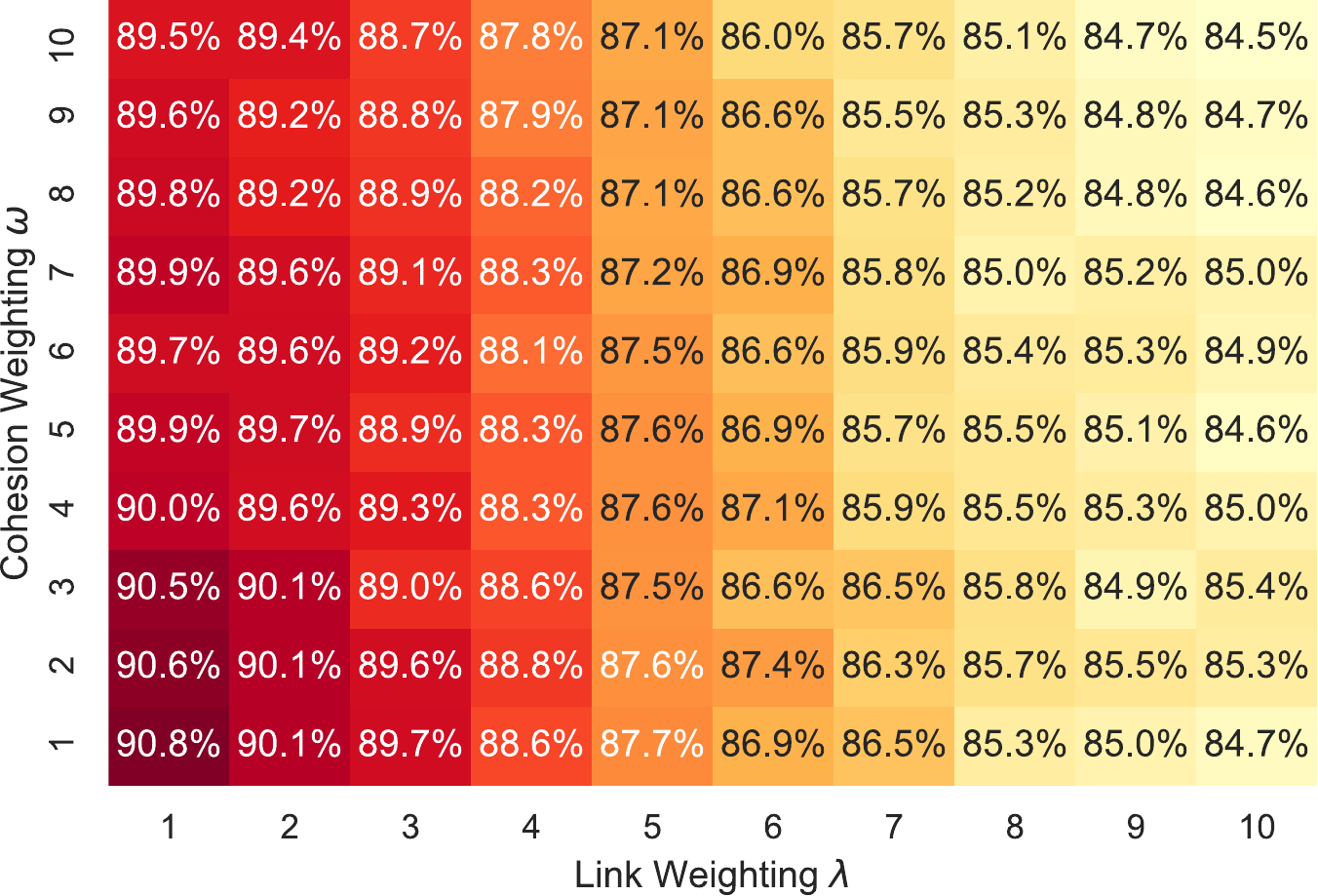}
	\vspace{-0cm}	
	\caption{Parameter evaluation for a rural \ac{REP}.}
	\label{fig:hm_friis}
	\vspace{-0cm}	
\end{figure}

\begin{figure*}[] 
	\centering
	\subfloat[]{\includegraphics[width=0.315\textwidth]{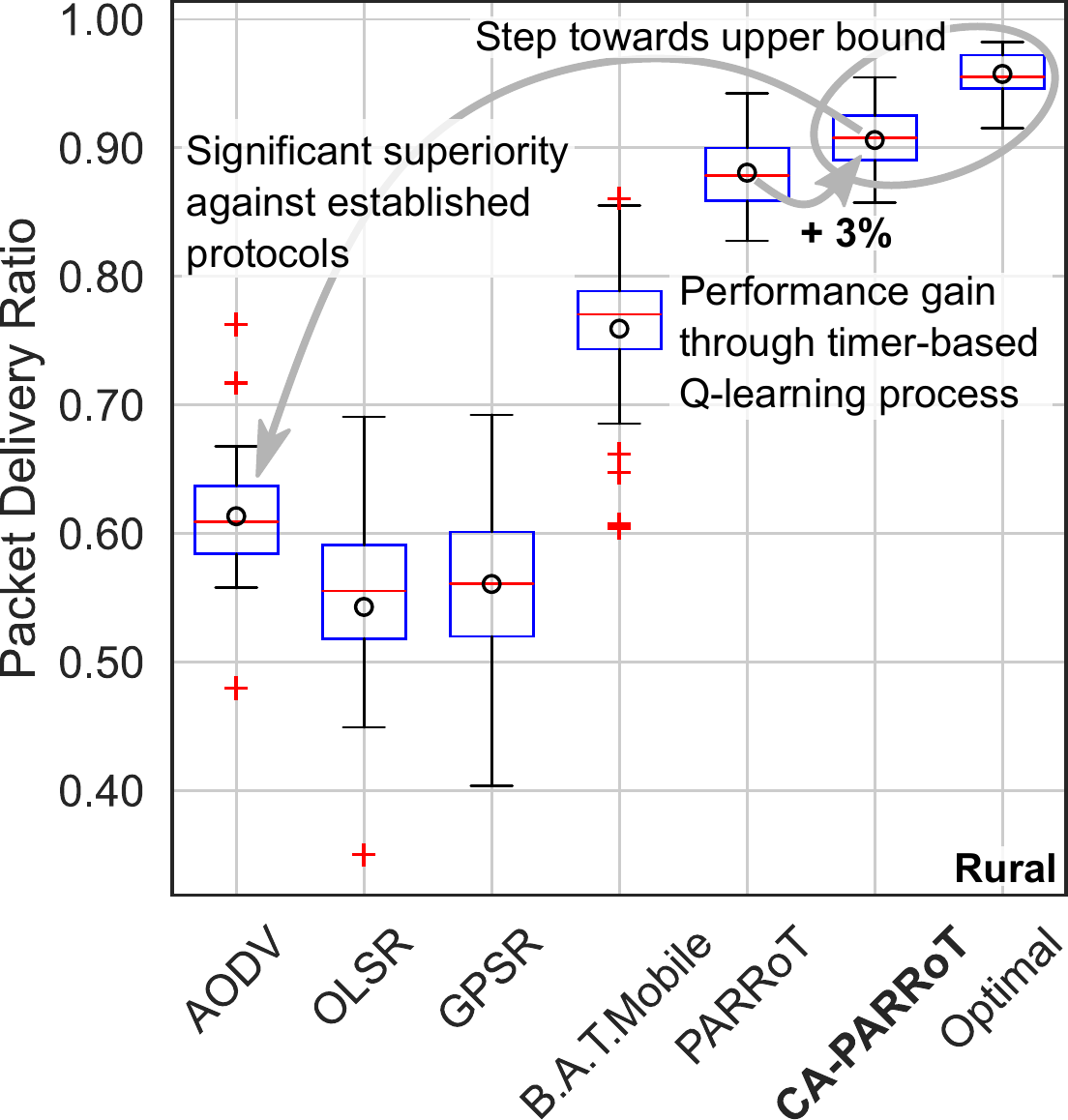}}\hfill
	\subfloat[]{\includegraphics[width=0.32\textwidth]{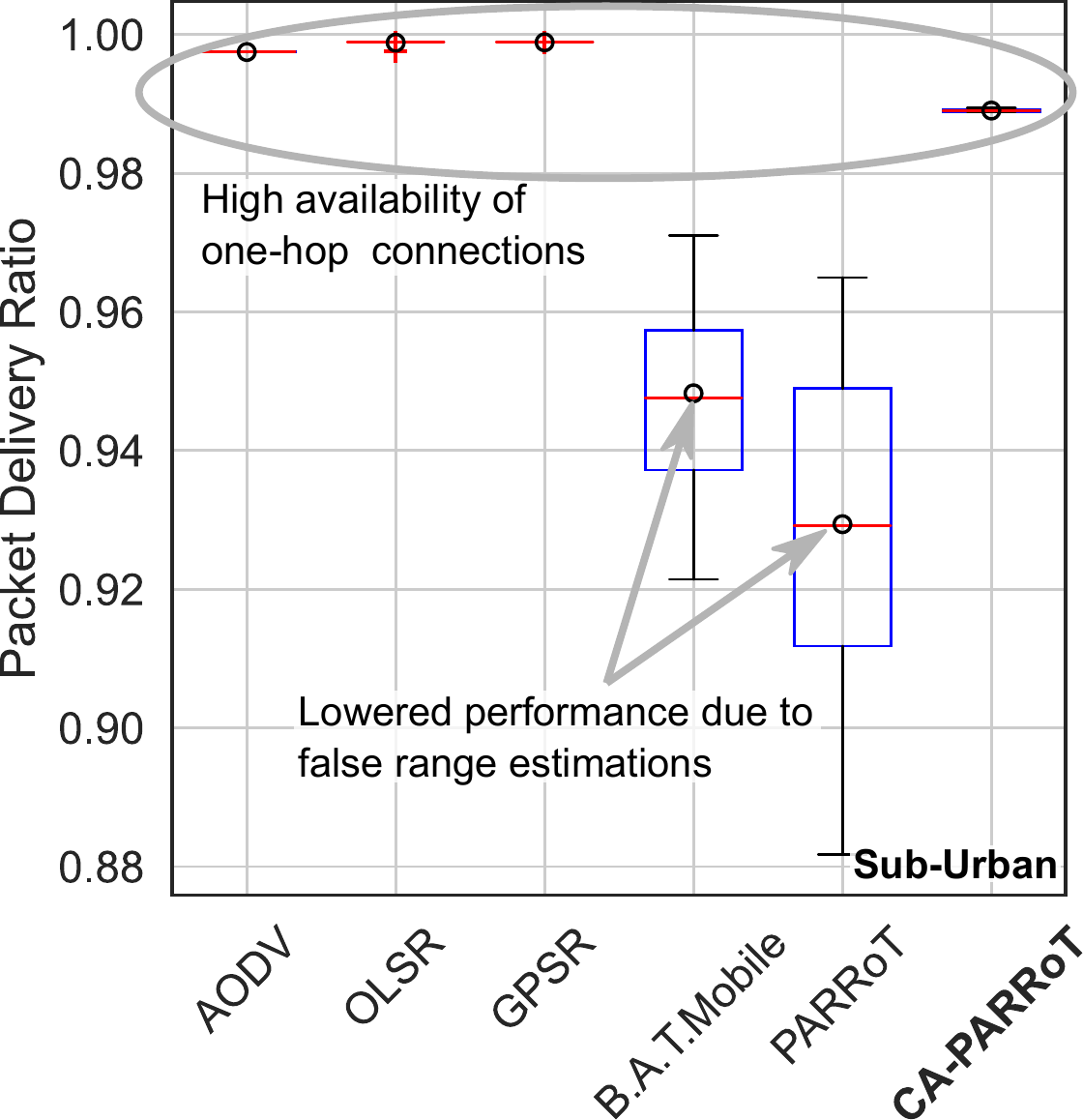}}\hfill
	\subfloat[]{\includegraphics[width=0.32\textwidth]{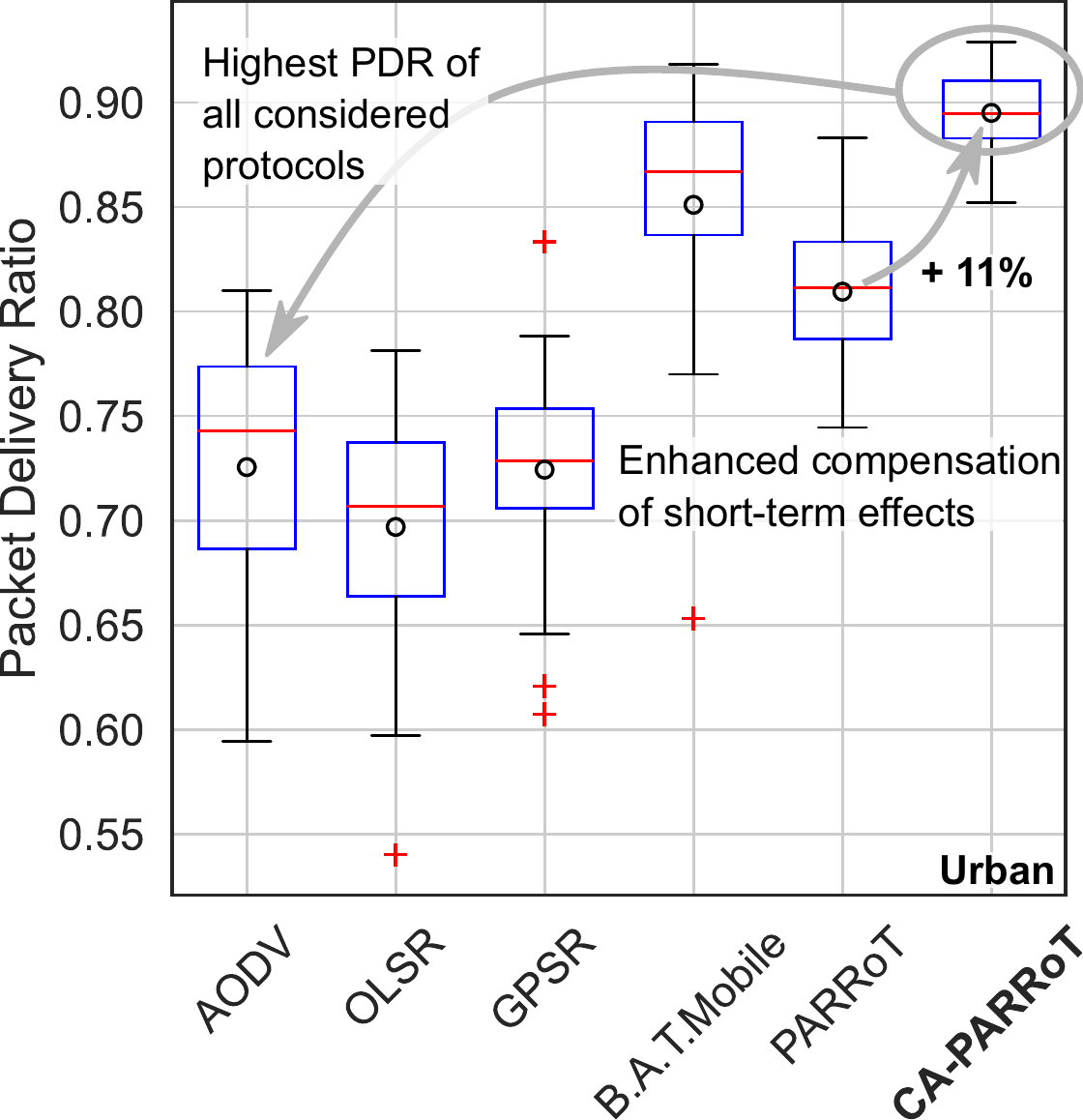}}\hfill	
	\caption{Comparison of the \ac{PDR} of different routing protocols across the rural (a), sub-urban (b), and urban (c) \acp{REP}.}
	\label{fig:pdr_comparison}
	\vspace{-0cm}	
\end{figure*}

\begin{figure*}[] 
	\centering
	\subfloat[]{\includegraphics[width=0.315\textwidth]{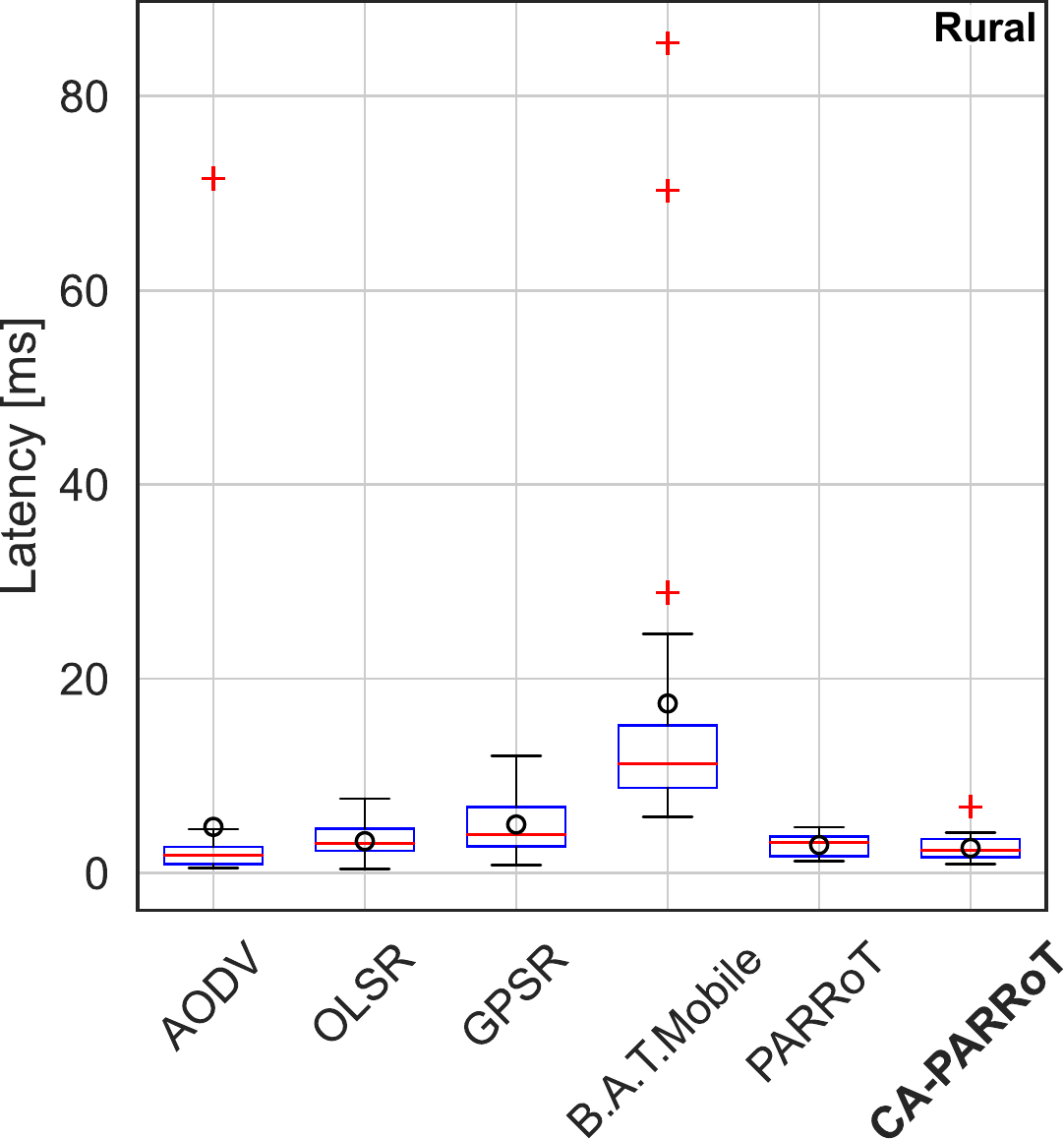}}\hfill
	\subfloat[]{\includegraphics[width=0.315\textwidth]{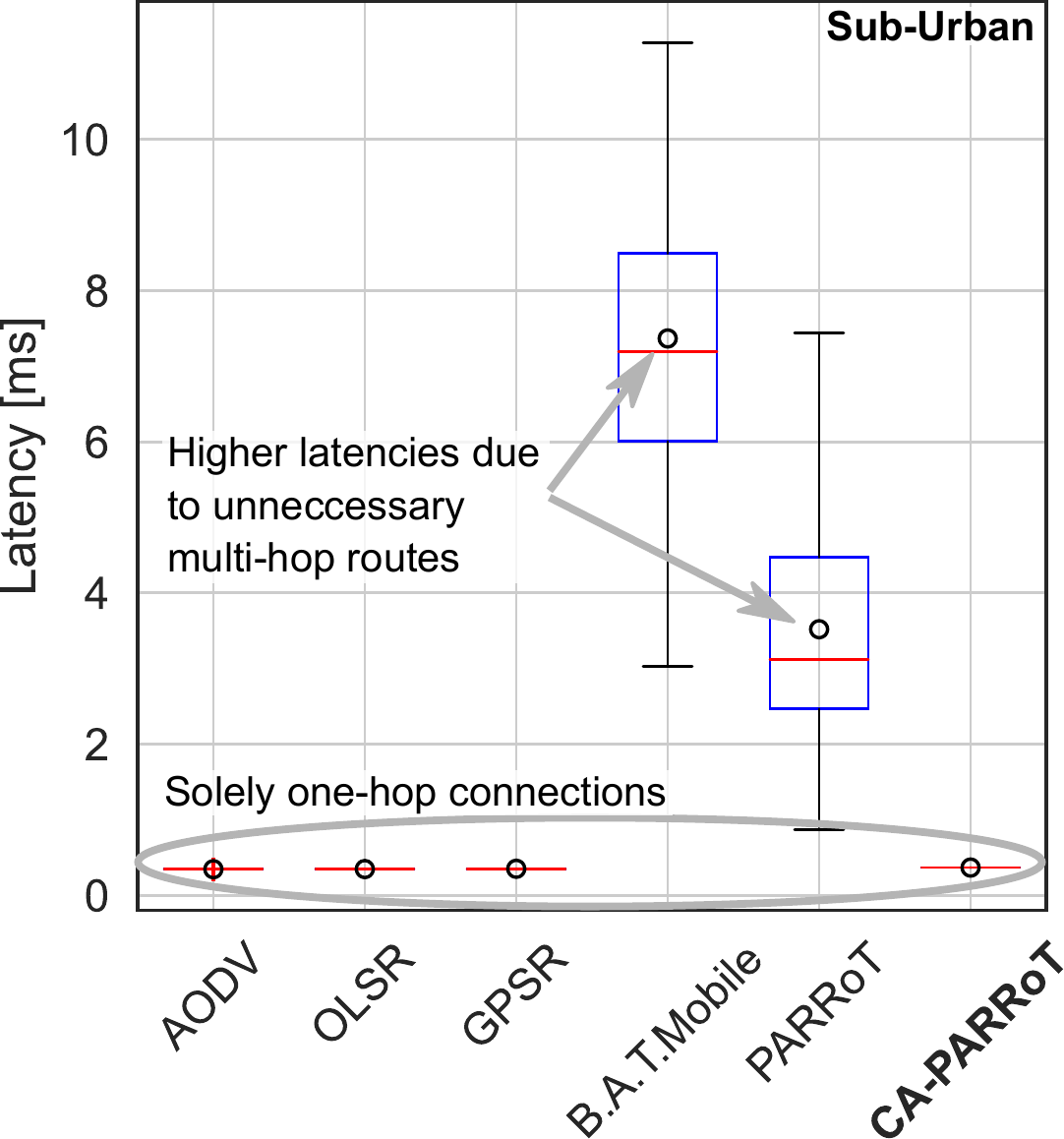}}\hfill
	\subfloat[]{\includegraphics[width=0.32\textwidth]{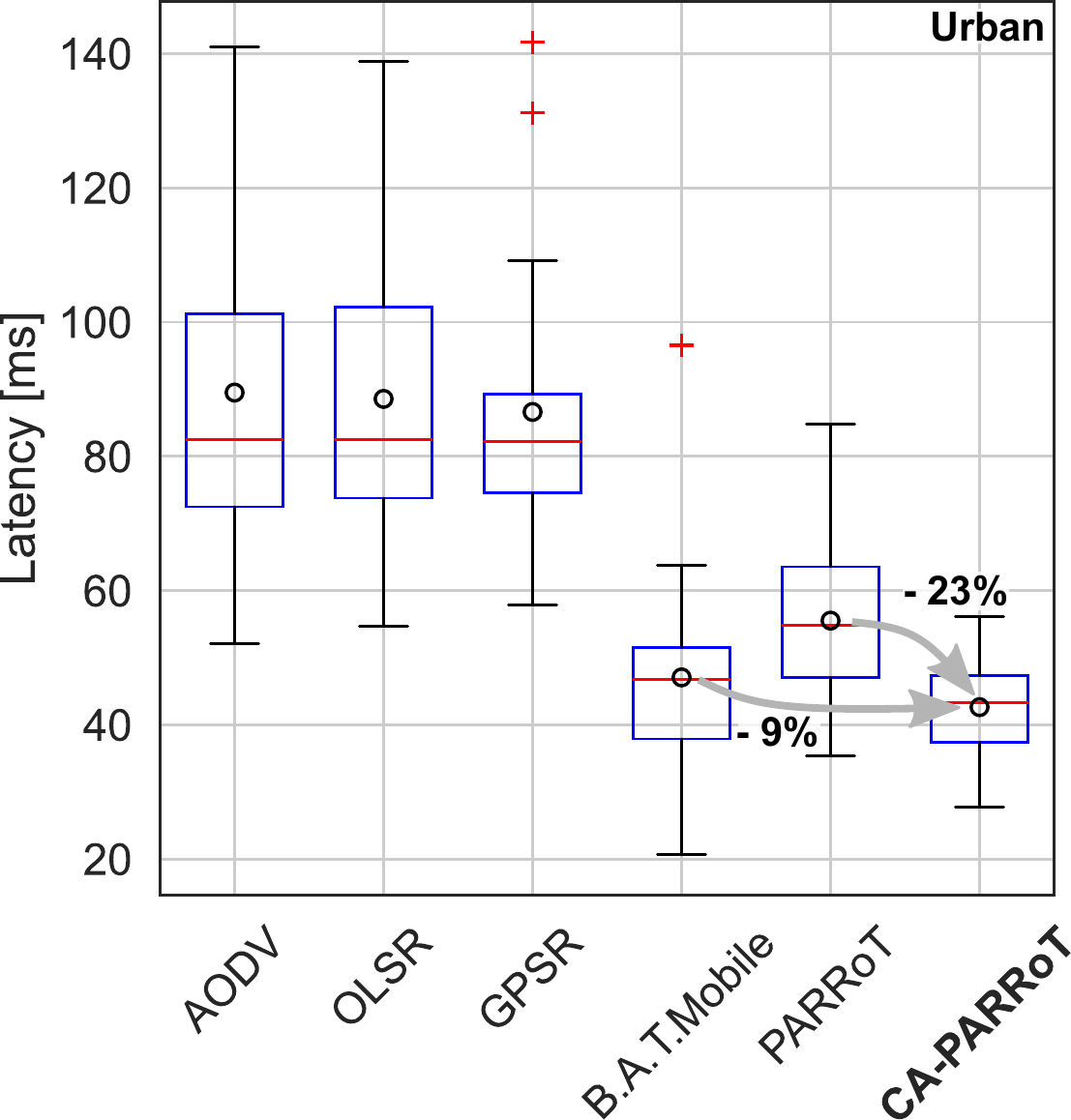}}\hfill	
	\caption{Comparison of the latency of different routing protocols across the rural~(a), sub-urban~(b), and urban~(c) \acp{REP}.}
	\label{fig:latency_comparison}
	\vspace{-0cm}	
\end{figure*}

\section{Results} \label{sec:results}
In this section, we present the evaluation of our proposal \ac{GULAG}.
First, the \ac{ML}-based \ac{REP} classification is elaborated with \textit{LIMITS}, which also offers the capability to export the models.
In parallel, we carry out a parameter study for each \ac{REP}, where we investigate the adjustment of the range estimation through the range budget $r_b$. 

The results are then merged in the new routing protocol \ac{GULAG}, whose end-to-end performance is evaluated and compared with other protocols in a live setup, where \ac{GULAG} is deployed without pre-parameterization and needs to prove its self adaptability.

\subsection{Configuration of \acl{GULAG}}
Fig.~\ref{fig:ml_accuracy} shows the cross validation accuracy for the considered \ac{ML} models \ac{ANN}, linear \ac{SVM}, and \ac{RF}. All models show a high accuracy and provide a reliable \ac{REP} classification from signal strength and distance pairs.
\acl{RF} shows to be the most accurate model and, thus, is chosen to be integrated into \ac{GULAG}.

The range budget $r_b$ is an important parameter, as it does not only impact the Q-learning, but also limits the expiration timeframe for all knowledge entries through the estimated \ac{LET}.
Fig.~\ref{fig:range_budget} shows the achieved \acp{KPI} for $r_b$ variations in urban and rural scenarios.
The rural prototype shows a massive performance degradation for positive range budgets, which correspond to an overestimation of the communication radius. In turn, a peak value for a slight underestimation can be observed, and, thus, a default of $-5$\,m for rural prototypes is determined.
Contrary, the urban prototypes shows an increasing latency for higher $r_b$, but also a higher \ac{PDR}. As a tradeoff conception, $20$\,m are selected as default, as the \ac{PDR} begins to plateau here, and the latency is lowest for this \ac{PDR} level.
The used two-ray ground model for sub-urban areas, is based on a smaller attenuation coefficient, which leads to a higher communication range. Thus, $r_b = 600$\,m is set, to cover the biggest possible distance within the scenario playground.

\begin{table}[ht]
	\centering
	\caption{Resulting Parameters for Different \acp{REP}}
	\begin{tabular}{lrrr}
	
	\toprule	
	\textbf{Parameter} & \textbf{Rural} & \textbf{Sub-urban} & \textbf{Urban} \\
	
	\midrule
	Range Budget $r_b$ & \SI{-5}{\meter} & \SI{600}{\meter} & \SI{20}{\meter} \\
	Learning Rate $\alpha$ & 0.5 & 0.2 & 0.6 \\
	Basic Discount $\gamma_0$ & 0.8 & 0.2 & 0.3 \\
	Link Weighting $\lambda$ & 1 & 3 & 1\\
	Cohesion Weighting $\omega$ & 1 & 2 & 2\\
	\bottomrule
	
	\end{tabular}
	\label{tab:param_db}
\end{table}
The metric weightings $\lambda$ and $\omega$ are evaluated for intervals of $[1, 10]$ each, which result in a heatmap as shown in Fig.~\ref{fig:hm_friis} for a rural \ac{REP}.
The parameter optimization for the \ac{RL} parameters is carried out according to \cite{Sliwa/etal/2021a}.
Tab.~\ref{tab:param_db} compromises the resulting parameterizations of \ac{PARRoT} for rural, sub-urban, and urban \acp{REP}.

\subsection{Performance Evaluation in a Three-Dimensional Random Waypoint Scenario}\label{sec:results:C:3D_RWP}
Fig.~\ref{fig:pdr_comparison} shows the results of the performance evaluation across different \acp{REP} for the considered reference protocols and the \ac{ML}-based \ac{GULAG}.
In the shown scenarios, \ac{GULAG} was deployed without manual configuration so that the \ac{ML} component was responsible to classify the \ac{REP} and select the appropiate parameter set.

For a rural \ac{REP}, a performance gain of $3\,\%$ compared to \ac{PARRoT} can be observed and, thus, only a gap of $5\,\%$ remains to the theoretical upper bound, which expresses the mobility-constrained availability of routes. B.A.T.Mobile is outperformed by $19\,\%$, and \ac{AODV}, as best performing established reference protocol, can be exceeded by $48\,\%$.
Also, considering the latency, \ac{GULAG}'s latency is $10\,\%$ lower than \ac{PARRoT}'s, and $21\,\%$ lower than \ac{OLSR}'s.

As described before, the channel model for the sub-urban \ac{REP} considers a higher communication range. Therefore, Fig.~\ref{fig:pdr_comparison}~(b) shows a high availability of one-hop connections across all protocols. B.A.T.Mobile and \ac{PARRoT} both use range estimations in their routing process, derived from a freespace model with higher attenuation, and, thus, being too low for this \ac{REP}. This causes both protocols to plan multi-hop routes although one-hop connections would be available.
Here, the better fitting parameter set of \ac{GULAG} with a higher range budget $r_b$ prevents the protocol from being too pessimistic, thus, also deciding to route packets directly to their destination.

Fig.~\ref{fig:pdr_comparison}~(c) and Fig.~\ref{fig:latency_comparison}~(c) show the achieved \acp{KPI} for the urban \ac{REP}, characterized by the highest channel variance of the considered \acp{REP}.
As shown, the proposed timer-based compensation technique and \ac{ML}-based classification component enable \ac{GULAG} to overcome the challenges, posed by the urban \ac{REP}, and raise the \ac{PDR} by $11\,\%$, while reducing latency by $23\,\%$ compared to our previous protocol \ac{PARRoT}.

Compared to B.A.T.Mobile, the former best performing protocol in \cite{Sliwa/etal/2021a}, \ac{GULAG} achieves a $5\,\%$ higher \ac{PDR} and a $9\,\%$ lower latency. The established reference protocols are outperformed by up to a $23\,\%$ better \ac{PDR} and $50\,\%$ lower latency.

\section{Conclusion}
In this paper, we presented \ac{GULAG} as an extension of \ac{PARRoT}, that uses a timer-based update process to compensate short-term effects in challenging surroundings.
The dedicated machine learning component for predicting the current \ac{REP} enables \ac{GULAG} to choose the best known parameter set and, thus, being able to operate autonomously in different \acp{REP}.
As shown in comprehensive simulations, \ac{GULAG} achieves a significant gain compared to \ac{PARRoT} and outperforms the considered other protocols.

In future work, we will consider more radio prototypes and evaluate type-dependent optimization techniques to be applied depending on the predicted \ac{REP}.

\ifdoubleblind

\else

	\section*{Acknowledgment}
	
	\footnotesize
	This work has been supported by the German Research Foundation (DFG) within the Collaborative Research Center SFB 876 ``Providing Information by Resource-Constrained Analysis'', projects A4 and B4 as well as by the Ministry of Economic Affairs, Innovation, Digitalization and Energy of the state of North Rhine–Westphalia in the course of the Competence Center 5G.NRW under grant number 005–01903–0047, and in the course of the project Plan \& Play under grant number 005-2008-0047.

\fi

\ifacm
\else
	\bibliographystyle{IEEEtran}
	\bibliography{Bibliography}
\fi

\clearpage

\end{document}